\documentclass[12pt]{iopart}
\usepackage{graphicx}
\usepackage[utf8]{inputenc}
\usepackage[T1]{fontenc}
\usepackage{braket}
\usepackage{textcomp}
\usepackage{SIunits}
\usepackage{hyperref}
\DeclareUnicodeCharacter{B5}{\ifmmode\mu\else\textmu\fi}
\usepackage[usenames, dvipsnames]{color}

\begin{document}
\title{A cryogenic radio-frequency ion trap for quantum logic spectroscopy of highly charged ions}

\author{T Leopold$^1$, S A King$^1$, P Micke$^{1,2}$, A Bautista-Salvador$^1$, J C Heip$^1$, C Ospelkaus$^{1,3}$, J R Crespo~L\'opez-Urrutia$^{2}$ and P O Schmidt$^{1,3}$}
%\homepage[]{Your web page}
%\thanks{}
%\altaffiliation{}
\address{$^1$Physikalisch-Technische Bundesanstalt, Bundesallee 100, 38116 Braunschweig, Germany}
\address{$^2$Max-Planck-Institut f\"ur Kernphysik, Saupfercheckweg 1, 69117 Heidelberg, Germany}
\address{$^3$Institut f\"ur Quantenoptik, Leibniz Universit\"at Hannover, Welfengarten 1, 30167 Hannover, Germany}

\ead{piet.schmidt@quantummetrology.de}
\vspace{10pt}
\begin{indented}
\item[]\today
\end{indented}

\begin{abstract}
A cryogenic radio-frequency ion trap system designed for quantum logic spectroscopy of highly charged ions is presented. It includes a segmented linear Paul trap, an in-vacuum imaging lens and a helical resonator. We demonstrate ground state cooling of all three modes of motion of a single $^9$Be$^+$ ion and determine their heating rates as well as excess axial micromotion. The trap shows one of the lowest levels of electric field noise published to date. We investigate the magnetic-field noise suppression in cryogenic shields made from segmented copper, the resulting magnetic field stability at the ion position and the resulting coherence time. Using this trap in conjunction with an electron beam ion trap and a deceleration beamline, we have been able to trap single highly charged Ar$^{13+}$ (Ar XIV) ions concurrently with single Be$^+$ ions, a key prerequisite for the first quantum logic spectroscopy of a highly charged ion.
\end{abstract}

\section{Introduction}\label{I}
Over the last decade, there has been growing interest in high precision spectroscopy of highly charged ions (HCI) for applications in frequency metrology and fundamental physics \cite{kozlov_highly_2018,safronova_search_2018}, such as the search for a possible variation of fundamental constants \cite{schiller_hydrogenlike_2007, berengut_enhanced_2010, berengut_optical_2012, safronova_highly_2014}, violation of local Lorentz invariance \cite{dzuba_strongly_2016}, or probing for new long-range interactions \cite{berengut_probing_2018}. The strong scaling of energy levels with charge state shifts fine and hyperfine transitions into the optical regime \cite{kozlov_highly_2018,gillaspy_highly_2001,berengut_optical_2012}, enabling high-precision laser spectroscopy. The highest sensitivity to many of the tests of fundamental physics can be found in optical transitions between levels of different electronic configuration near energy-level crossings as a function of charge state \cite{berengut_highly_2012}. The small size of electron orbitals in HCI and correspondingly reduced atomic polarisability and electric quadrupole moment, suppresses field-induced systematic frequency shifts, suggesting HCI as favourable optical clock candidates \cite{schiller_hydrogenlike_2007,derevianko_highly_2012,berengut_optical_2012,berengut_testing_2013,safronova_highly_2014-1,dzuba_optical_2015,yu_selected_2018}.\par 

Up to now, precision spectroscopy of HCI at  rest was mostly performed in electron beam ion traps (EBITs). However, high ion temperatures ($T>10^5$\,K) due to the electron impact heating in a deep trapping potential and magnetic field inhomogeneities as well as drifts have limited the achievable spectroscopic resolution and accuracy in most cases to the parts-per-million level. Recently, the transfer of HCI from an EBIT to a Paul trap and sympathetic cooling to millikelvin temperatures using co-trapped laser-cooled Coulomb crystals was demonstrated \cite{schmoger_coulomb_2015-1}. Combined with ultrastable local oscillator technology \cite{kessler_sub-40-mhz-linewidth_2012, matei_1.5_2017} and the techniques used in optical frequency standards based on quantum logic \cite{schmidt_spectroscopy_2005, ludlow_optical_2015}, this paves the way for a 10$^9$ to 10$^{13}$-fold improvement over the current most accurate spectroscopic measurements for HCI \cite{draganic_high_2003,bekker_2018}.\par 

Currently there are only two approaches for high-precision optical spectroscopy of HCI. One is the use of Penning traps, where a single HCI is trapped and resistively cooled to the temperature of the cryogenic ion trap \cite{brantjes2011penning,sturm_priv}. Successful laser-induced excitation of the transition is detected by electronic readout of the ion spin by coupling it to its motion in an inhomogeneous magnetic field \cite{haffner_double_2003}. The other approach, followed by this experiment, is based on the detection of Rabi flopping on the spectroscopy HCI using a co-trapped singly-charged ion such as $^9$Be$^+$, which also provides sympathetic cooling. Optical excitation of the HCI is transferred to the Be$^+$ ion exploiting the coupled motion of the ions using well-established quantum logic protocols \cite{schmidt_spectroscopy_2005}. Then, the excitation of the Be$^+$ ion is read out by fluorescence detection.\par 
The HCI lifetime in a trap is limited by charge-exchange reactions with background gas particles. For frequency metrology, useful lifetimes in the minutes to hours regime can only be achieved in cryogenic environments with pressure levels of typically below $10^{-14}$~mbar \cite{schwarz_cryogenic_2012,repp_pentatrap:_2012,pagano_cryogenic_2019}. 
Here, a cryogenic ion trap setup including a newly designed segmented blade trap, an in-vacuum helical resonator and a imaging system with high collection efficiency is presented. The segmented blade trap is characterised in terms of ion heating rates and excess axial micromotion. Furthermore, ground state cooling of all three normal modes of motion of a single Be$^+$ ion confined far outside the Lamb-Dicke regime is demonstrated. The magnetic shielding due to the cryogenic heat shields and the corresponding passive stability of the magnetic field is evaluated. We report on measurements of the coherence time and suppression of low-frequency magnetic field noise. Finally, we demonstrate the successful loading and storage of a single ion of Ar$^{13+}$ (Ar XIV).\par 
The entire system is inspired by \cite{schwarz_cryogenic_2012} and has consequently been named "Cryogenic Paul Trap Experiment PTB" (CryPTEx PTB). Compared to \cite{schwarz_cryogenic_2012} our much smaller trap size yields higher secular frequencies, which are required for high-fidelity quantum logic operations \cite{wineland_experimental_1998}. Furthermore, the trap is connected to a low-vibration cryogenic supply line \cite{micke_cryosupply_2018} and a compact EBIT producing the desired HCI \cite{micke_heidelberg_2018}. \par

\section{Apparatus}\label{apparatus}
\begin{figure*}
	\centering
	\includegraphics[width=0.8\linewidth]{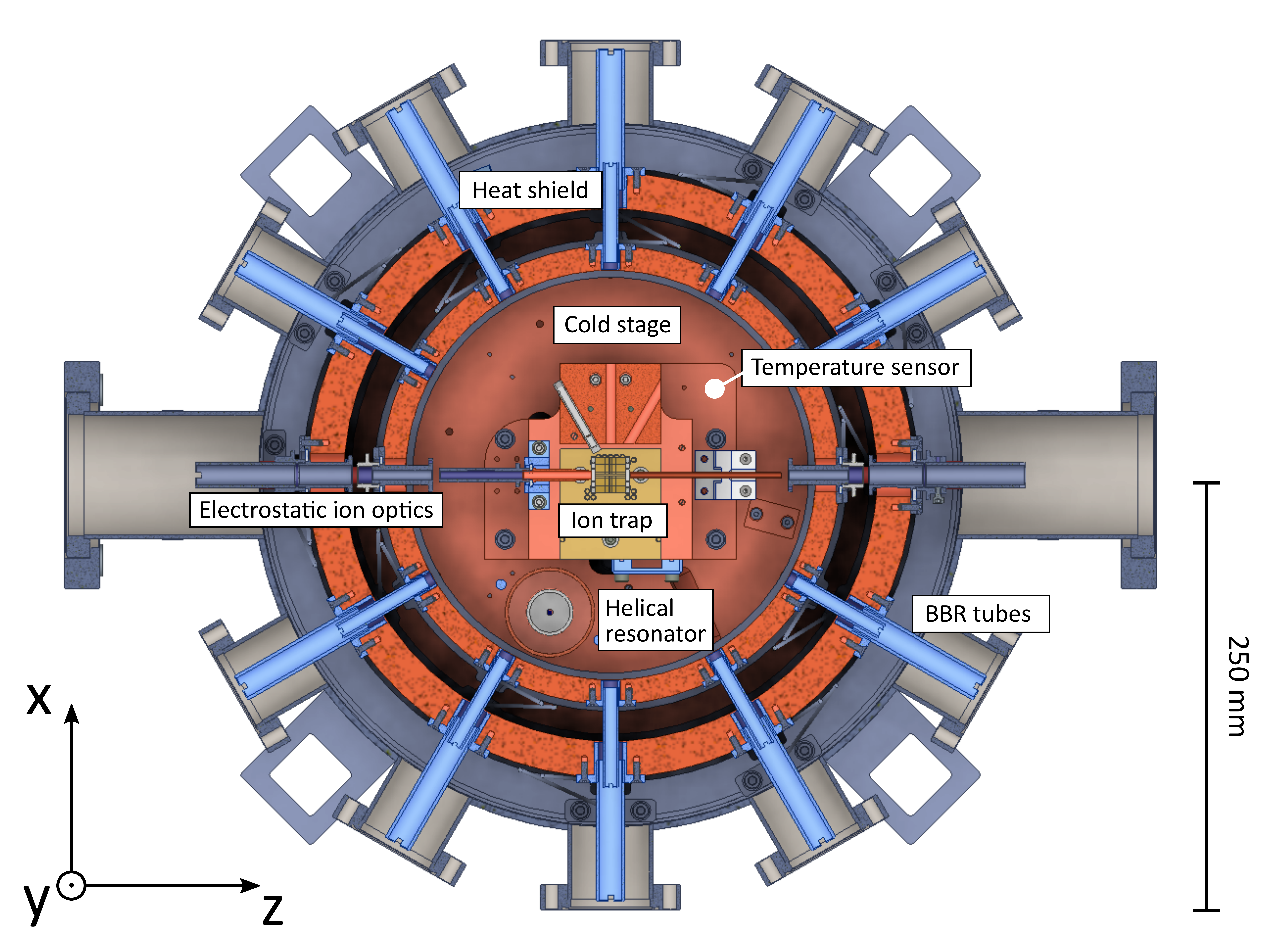}% % Important NOTE: Please make certain your figures do not include local directory paths. ex. "c:\file\sub\fig1.eps"
	\caption{Top view cross section of the CryPTEx PTB apparatus described in the text. The in-vacuum imaging system is not shown. 
		\label{fig:setup}}%
\end{figure*}
Cryogenic quadrupole-Paul traps already exist in several laboratories around the world, mainly as surface-electrode ion traps \cite{labaziewicz_suppression_2008,antohi_cryogenic_2009,wang_superconducting_2010,niedermayr_cryogenic_2014,vittorini_modular_2013}. There, a suppression of the anomalous heating rates due to reduced electric field noise is exploited \cite{deslauriers_scaling_2006,labaziewicz_suppression_2008}. Macroscopic cryogenic Paul traps are in use for reaching extremely low pressures\cite{schwarz_cryogenic_2012, pagano_cryogenic_2019}, serving as a reservoir for buffer-gas cooling of molecules \cite{hansen_efficient_2014}, or reducing black-body radiation shifts in optical clocks \cite{rafac_sub-dekahertz_2000}. The setup presented here is the first cryogenic Paul trap designed for sub-Hz-level spectroscopy of optical transitions in HCI. \par 
The lifetime of trapped HCI depends on the mean free time between collisions with the background gas. Due to the keV-level ionisation potential of HCI, collisions of that type will likely result in charge exchange between the HCI and the neutral partner. This can cause HCI ejection from the trap, which typically has a depth of only a few eV/charge in the axial direction. Even if the ion is retained after the charge-exchange event, it does not longer belong to the required spectroscopic species, i.~e., charge state, and must be replaced. With cryopumping, pressures of less than $10^{-16}$~mbar can be achieved \cite{diederich_observing_1998}. As a comparison, most atomic physics experiments operate at pressures of $10^{-11}$~mbar or above at room temperature.\par 

\subsection{Cryogenic setup}
In addition to the usual requirements, the design of a cryogenic ion trap for frequency metrology calls for:
\begin{enumerate}
	\item Extremely stable cold trap mounting inside the room-temperature vacuum chamber, while thermally insulating  the cold stages;\\
	\item Multiple optical ports for lasers, imaging, and external ion delivery while keeping blackbody radiation (BBR) heat load from outside as low as possible;\\
	\item Electrical connections, including the radio-frequency drive, should use long cables with very low thermal conductivity;\\ 
	\item Repeated thermal cycling needed without trap damage or misalignment relative to the external optical setup;\\
	\item Vibrations generated at the liquid-helium cryostat or the mechanical cryocooler need to be suppressed at the position of the ion trap.
\end{enumerate}
The ion trap setup is attached to a closed-cycle low-vibration cryogenic system, reaching temperatures of $<5$~K and 50~K on the cold stage and heat shield, respectively, despite the $\sim$~1.4~m separation from the cold head \cite{micke_cryosupply_2018}. It suppresses the pulse tube vibrations to values below 20~nm in the horizontal plane spanned by the laser beams and below 100~nm in the vertical direction. Figure \ref{fig:setup} shows a CAD cross-sectional view of the setup. It consists of a nested structure of temperature stages, each mounted symmetrically to the previous one using $\sim$~15~cm-long stainless-steel spokes to minimise heat conduction between them in spite of the rigid mechanical connection. The vacuum chamber is fixed on the optical table, where the laser setup rests. Both the heat shield and the cold stage consist of a base plate, a wide tube and a lid made of high-purity (99.99~\%), gold-plated oxygen-free high conductivity (OFHC) copper. Gold plating both enhances the thermal conductivity at contact points and prevents tarnishing of the copper during periods when the system is vented, thus maintaining a low emissivity in order to reduce heat transfer via BBR. A wall thickness of 15~mm on the heat shield and 10~mm on the cold stage efficiently shield external AC electromagnetic fields, and particularly well at low temperatures, where the electrical conductivity of the copper is two orders of magnitude higher than at room temperature. Convenient access to the inside of the cold stage is provided by just removing three lids, allowing maintenance work without detaching the cryogenic setup.\par
Symmetric arrangement of the stainless steel spokes, as in reference \cite{schwarz_cryogenic_2012,micke_cryosupply_2018}, minimises displacements during thermal cycling. After a complete thermal cycle, the trapped ions can be optically addressed without realignment of the lasers or imaging system. Optical access is provided by 16 ports, 12 of which are equally spaced in the horizontal plane. Four additional ports are provided at 15$^\circ$ off the horizontal plane, spanning a vertical plane with the trap axis. The ports restrict the solid angle using two nested concentric, 5~cm-long aluminium tubes on the cold stage and heat shield with an inner diameter of 5~mm and 11~mm respectively. This minimises both the exposure of the cold stage to room-temperature BBR and the flux of residual-gas particles into the cold stage, thereby enhancing differential pumping from the room-temperature sections. The solid angle fraction of room-temperature elements visible to the ion is only 0.01~\%, which improves upon the $\sim 2~\%$ reported previously \cite{schwarz_cryogenic_2012}.\par
All necessary DC electrical connections are provided by means of 2~m-long phosphor-bronze wires with a diameter of 200~µm (Lakeshore QT-32), thermally anchored at the heat shield and cold stage with 1~m of wire between the different temperature stages to reduce the thermal flow to the trap electrodes. For the trap drive radio-frequency (RF) signal and a microwave antenna, designed to drive the 1.25~GHz Be$^+$ ground state hyperfine structure splitting, we use semi-rigid coaxial beryllium-copper wire (Coax Co., SC-219/50-SB-B).\par 
The estimated heat load of the ion trap setup onto the cryostat is given in table \ref{tab:heat_load}. It is apparent, that the heat load on the cold stage is dominated by room-temperature BBR, while the heat load onto the cold stage is dominated by the dissipated RF power of the ion trap and thermal conduction through mechanical and electrical connections.
\begin{table}[htb]
	\centering
	\begin{tabular}{lcc}
		\hline
		\hline
		Heating mechanism & Heat load on heat shield (mW) & Heat load on cold stage (mW) \\
		\hline 
		Conduction through spokes & 780 & 38 \\
		Conduction through wires & 65 & 7 \\
		Trap drive RF & - & 200 \\
		Black-body radiation & 4000 & 19 \\
		\hline 
		Total & 4845 & 264 \\
		\hline
		\hline
	\end{tabular}
	\caption{Estimated steady-state heat load for temperatures of 50~K and 4.5~K on the heat shield and cold stage, respectively. \label{tab:heat_load}}
\end{table}

\subsection{The ion trap}
Our present design meets the requirements for quantum logic spectroscopy of HCI. The most important design goals were:
\begin{enumerate}
	\item low differential contraction between parts when cooling the trap to 4~K;
	\item wide axial access and large trap aperture for efficient injection of HCI from the deceleration beamline;
	\item a long axial trap for confining Coulomb crystals of several hundred Be$^+$ ions capable of efficiently stopping the HCI after injection;
	\item small RF electrode separations causing high secular frequencies needed for quantum logic spectroscopy.
\end{enumerate}
To meet the first criterium, the trap consists of alumina (sintered Al$_2$O$_3$), with adequate thermal conductivity at both room and cryogenic temperatures of 30 and 0.3 W/(m$\cdot$K) \cite{simon_cryogenic_1994}, respectively, and a low RF loss tangent of $2\cdot 10^{-4}$ \cite{dolezal_analysis_2015}. Our geometry is inspired by a segmented blade design \cite{gulde_experimental_2003}, however, the end caps are removed from the axis \cite{rowe_transport_2002,schaetz_towards_2007,hemmerling_single_2011}. Instead, the DC blades are segmented into five separate electrodes, whereby radially opposing electrodes can be biased to provide axial confinement. This breaks the cylindrical symmetry of the trap, lifting the degeneracy of the radial secular motional frequencies into a focused and de-focused mode and thus strongly defining the principal radial axes of the trap to be along the blade axes. This ensures efficient cooling of all ion motional modes by laser beams at oblique angles in the horizontal plane. A 0.9~mm ion-electrode distance guarantees sufficient axial access for HCI injection while taking advantage of the strong inverse scaling of the anomalous ion-heating rate on the ion-electrode characteristic distance $d$ (between $d^{-2}$ and $d^{-4}$, depending on the nature of the noise source \cite{sedlacek_distance_2018}).\par
To avoid a direct line-of-sight from the ion to the insulator separating the DC electrodes, small slits with a width and depth of 100~µm are first cut into the blades at the desired electrode boundaries using a dicing saw. The electrode surfaces carry a 20~µm-thick gold layer. At first, a 20~nm-thick titanium seed layer and a 100~nm-thick gold layer were deposited on the alumina substrate using the multi-directional evaporation technique at the PTB surface-technology laboratory. Thereafter, a 20~µm-thick layer of gold was galvanically grown at the PTB clean-room facilities. This thickness was chosen to exceed the skin depth of the RF frequency drive (15~\micro m at 24~MHz at room temperature, decreasing to 1.5~\micro m at 4~K for a residual-resistivity ratio (RRR) of 100). The gold layer thickness was measured before and after the galvanic gold deposition using an optical microscope. After that, gold standing between the individual electrodes on the DC blade was removed by femtosecond-laser cutting at an external company (Micreon GmbH). Slits were also cut into the RF-carrying blades in the same pattern as in the DC blades to minimize axial micromotion \cite{herschbach_linear_2012}, however they were not laser-structured.\par 
\begin{figure}
	\centering
	\includegraphics[width=0.7\linewidth]{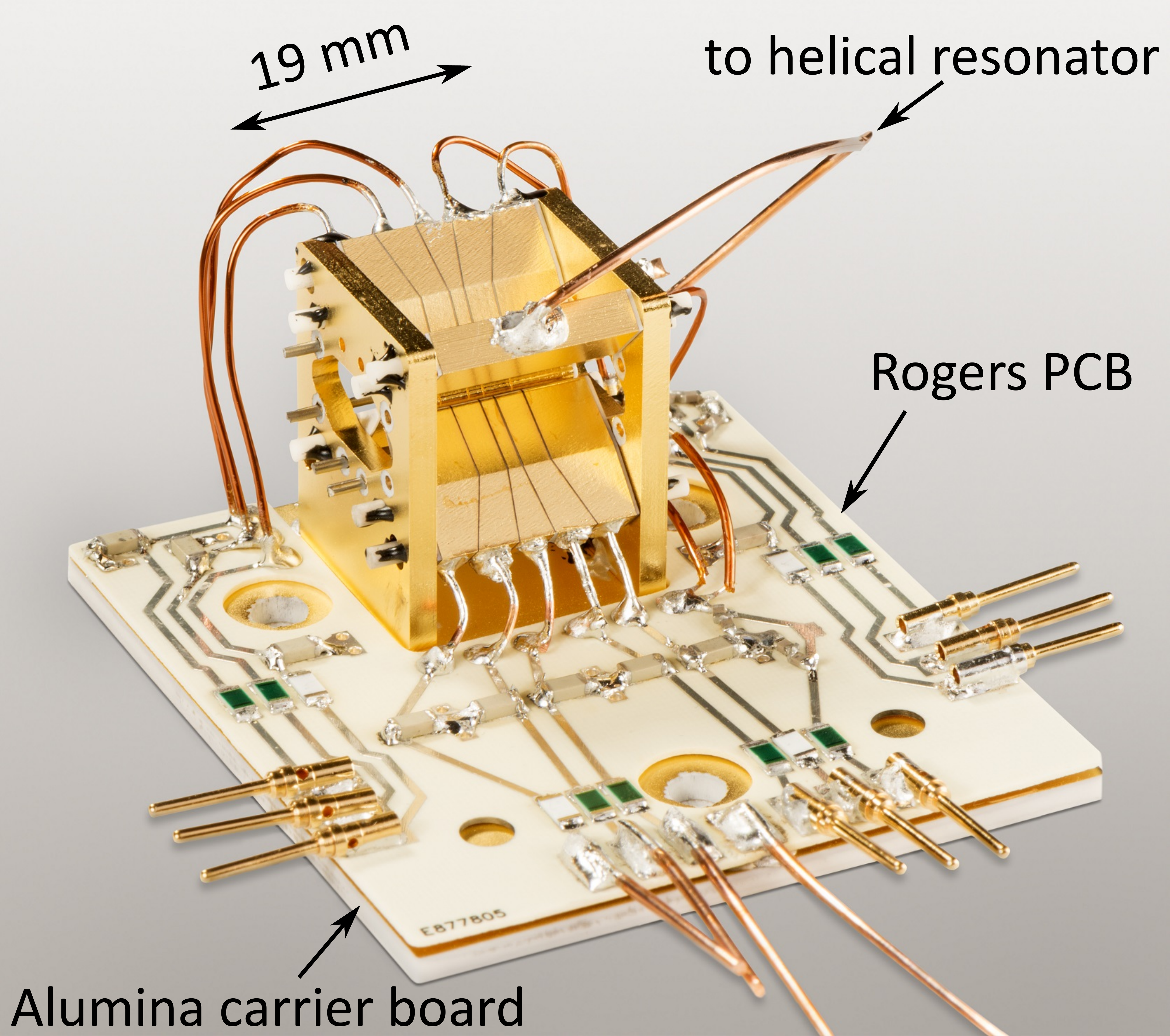}% % Important NOTE: Please make certain your figures do not include local directory paths. ex. "c:\file\sub\fig1.eps"
	\caption{Fully assembled ion trap with carrier and filter board. The gold-coated top layer of the carrier board (electrical ground of the trap) is electrically isolated from the vacuum chamber and can be biased to several 100 volts for HCI deceleration. \label{fig:assembled_trap}}%
\end{figure}
The fully assembled trap is mounted on a gold-coated alumina carrier board, which provides the RF and DC electrical ground. The carrier board also features a Rogers 4350B printed-circuit board (PCB) with surface-mounted device (SMD) filter elements and solder patches to connect to the DC blades, see figure \ref{fig:assembled_trap}. This provides a short path between filters and electrodes, reducing noise pickup. Additionally, mounting the filter components on the cold stage greatly reduces their Johnson noise.
\begin{figure}
	\centering
	\includegraphics[width=0.9\linewidth]{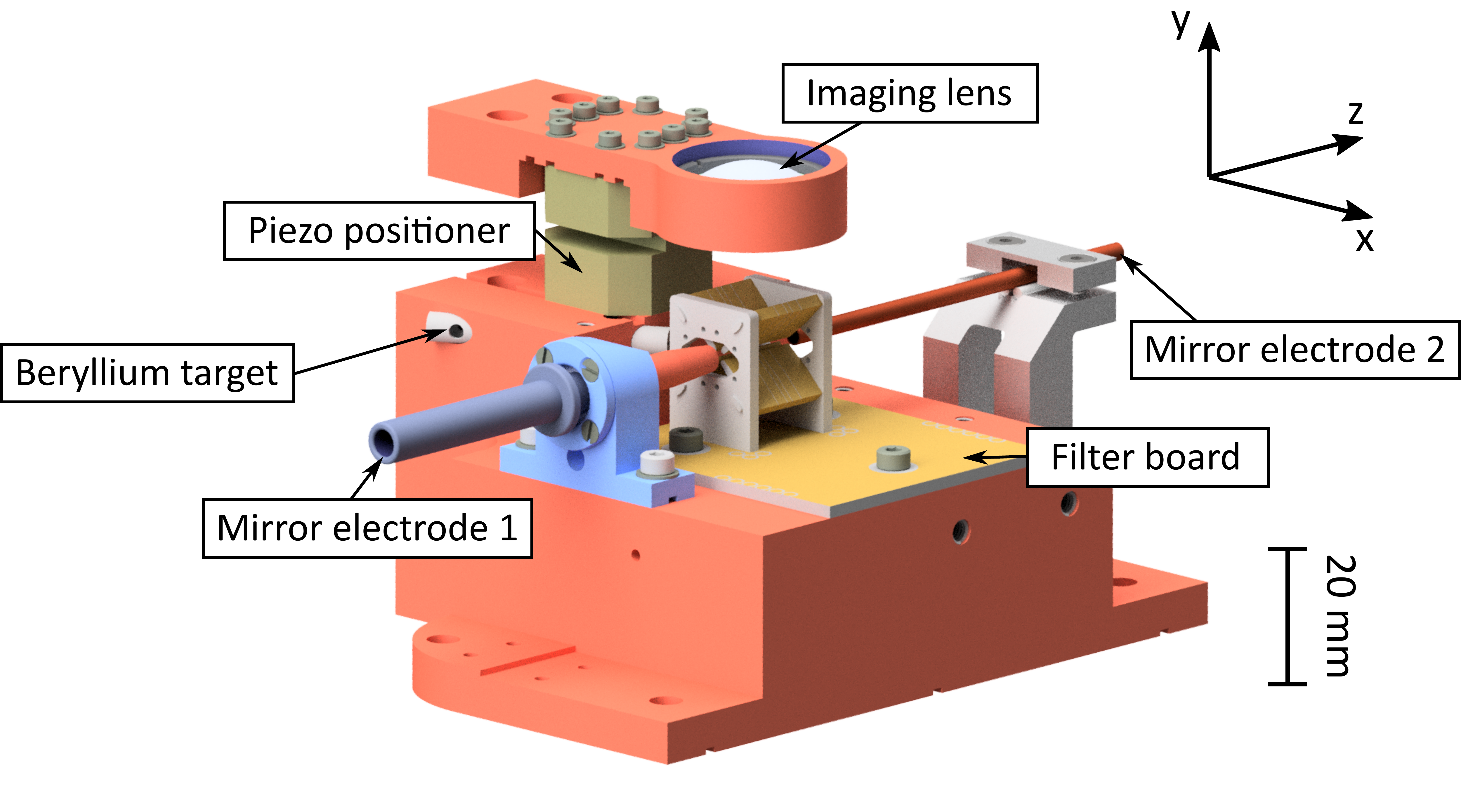}% % Important NOTE: Please make certain your figures do not include local directory paths. ex. "c:\file\sub\fig1.eps"
	\caption{CAD rendering of the ion trap assembly showing the in-vacuum imaging lens and electrostatic optics around the trap. \label{fig:trap_rendering}}%
\end{figure}
Connections from this filter board to the trap electrodes were made of Kapton-insulated copper wire, soldered at both ends using UHV compatible solder. The dielectric material of the SMD capacitors (Vishay VJ1206A472FF) is class C0G, proven to work at cryogenic temperatures with a sub-percent change in capacitance from room temperature \cite{teyssandier2010commercially}. All components were selected to be non-magnetic, including the Rogers PCB, which uses pure silver layers as conductor instead of the standard copper-nickel-gold combination. On the 6 inner DC blade electrodes and the compensation electrodes, the resistors (1~M$\Omega$, Vishay PNM1206E1004BST) and capacitors (4.7~nF) were selected to build a single-stage low-pass filter with a cutoff frequency of 34~Hz. For the outermost blade electrodes 1~k$\Omega$ resistors of the same series were chosen, increasing the cutoff frequency to 34~kHz to enable faster voltage switching whilst maintaining a low impedance for the RF pickup.\par
\begin{figure}	
	\centering
	\includegraphics[width=1\textwidth]{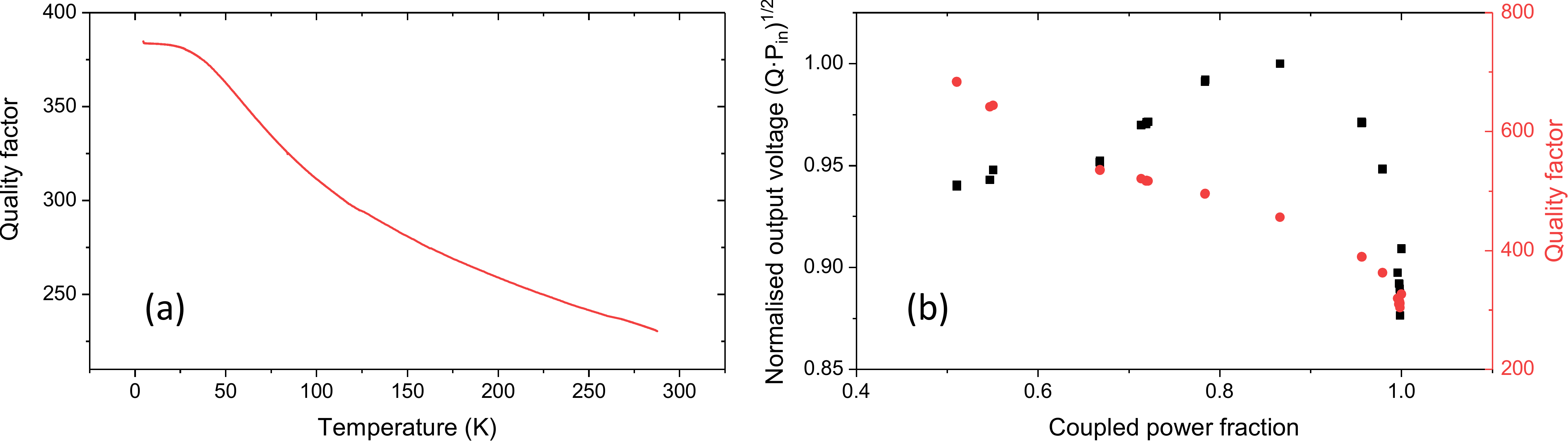} 
	\caption{(a) Helical resonator loaded quality factor $Q$ versus temperature. (b) Output voltage and $Q$ of the unloaded resonator as a function of the coupled power. The output voltage is optimised for a given level of power dissipation when there is a slight impedance mismatch. \label{fig:resonator}}
\end{figure}

 
The trap assembly is mounted on an OFHC copper block together with the imaging lens and ion-optic electrodes, see figure \ref{fig:trap_rendering}. The ion trap is driven by a miniature OFHC copper helical resonator placed next to the trap on the cold stage. The resonator coil is inductively coupled to a primary coil connected to an amplified low-noise signal generator (Rhode \& Schwarz SML01) using a resistive beryllium-copper coaxial cable and an insulated SMA connector feedthrough. For the resonator design we follow reference \cite{siverns_application_2012} for a room-temperature, unloaded quality factor $Q\approx 700$. It provides an unloaded resonance frequency near 50~MHz, and after loading with the trap electrodes 24.1~MHz. We monitor the value of $Q$ as a function of temperature during a thermal cycle using a network analyser: At room temperature, $Q=230$, and increases to $Q=385$ when cooled to 4.5~K as shown in figure \ref{fig:resonator}. This increase is much smaller than predicted based on to the RRR values of copper and gold, and shows a saturation behaviour starting at a temperature of around 50~K. This could be due to losses in the solder joints or the PTFE former around which the secondary coil is wound. \par 
By purposefully degrading the impedance matching between the primary and secondary coils, the mutual inductance of the resonator coils is reduced \cite{panja_note:_2015}, increasing $Q$. The voltage on the secondary coil scales with $\sqrt{Q\cdot P_{\textrm{in}}}$, where $P_{\textrm{in}}$ is the coupled power. Figure \ref{fig:resonator} (b) shows $Q$ and the calculated resonator output voltage as a function of $P_\textrm{in}$. We find that the output voltage for a given power dissipated in the resonator exhibits a maximum at slightly below 90~\% coupling. This is a relevant figure of merit in cryogenic systems only, where the dissipated power in the resonator can be the dominant heat load for the cold stage. Additionally, higher $Q$ leads to more effective filtering of noise on the trap- drive signal, reducing ion heating through parametric excitation \cite{wineland_et_al._experimental_1998,sedlacek_method_2018}.\par

For the final deceleration step of the HCI before they are loaded into the trap, we bias the whole trap to +200~V. This potential is added to the signal wire of the beryllium-copper coaxial cable using a bias-tee, keeping the shield of the helical resonator electrically isolated from the cold stage. A capacitor insulates the shield of the coaxial cable from the primary coil of the helical resonator. All other voltages required for ion trapping are referenced to this bias potential, which can be adjusted without affecting the trap operation.

\subsection{Loading and cooling of $^9$Be$^+$}
$^9$Be$^+$ ions are loaded into the trap by pulsed laser ablation and two-step photoionisation with a resonant intermediate level \cite{kjaergaard_isotope_2000}. A Q-switched frequency-doubled Nd:YAG laser at 532~nm with a pulse length of 4~ns and energy of up to 10~mJ is focused to a $1/e^2$ waist of 120~\micro m onto a beryllium wire located 18~mm from the trap centre. Ablated beryllium atoms are ionised using a 235~nm laser tuned to the $^{1}S_0$ – $^{1}P_1$ transition \cite{lo_all-solid-state_2014}. Typically, two laser pulses separated by 0.5~s are used for ablation, with a peak laser energy density of approximately 1.2~J/cm$^2$ for single ion loading and 4~J/cm$^2$ for loading of tens of ions, corresponding to peak intensities of around 300~MW/cm$^2$ and 1000~MW/cm$^2$ respectively.\par
For secular motional frequencies of a single $^9$Be$^+$ ion of 2.5~MHz in the radial direction and 1~MHz in the axial direction, 8~V have to be applied to the trap endcaps at an RF power of 200~mW. The RF-power dissipation raises the temperature of the cold stage by 0.3~K. \par
The ions are Doppler cooled on the strong, cycling $^{2}S_{1/2}$ $(F = 2)$ to $^{2}P_{3/2}$ $(F = 3)$ transition using a laser with a wavelength of 313~nm \cite{wilson_750-mw_2011,lo_all-solid-state_2014}. To lift the degeneracy of the Zeeman sublevels and define the quantisation axis, a bias magnetic field of 160~\micro T is applied at an angle of 30° to the trap axial direction in the horizontal plane using three mutually orthogonal pairs of coils. The cooling laser is delivered along the magnetic bias field axis with circular polarisation, as the cooling transition is closed for pure $\sigma$-polarised light. Residual polarisation impurities optically pump the ion to the $^{2}S_{1/2}$ $(F = 1)$ state, from which they are repumped with a separate laser tuned to the $^{2}S_{1/2}$ $(F = 1)$ to $^{2}P_{1/2}$ $(F = 2)$ transition. This beam is collinear to and has the same polarisation as the cooling laser.\par
A third 313~nm laser with a red detuning of 103~GHz to the cooling laser allows driving stimulated Raman transitions between the ground state hyperfine levels, used for ground-state cooling using resolved sidebands \cite{monroe_resolved-sideband_1995,wan_efficient_2015} and quantum logic operations \cite{schmidt_spectroscopy_2005}. A set of three Raman beams enables cooling and logic operations with projection purely onto either the axial or radial direction as necessary. The linear polarisation of each of the Raman beams was carefully tuned in order to minimise associated Stark shifts, albeit at the cost of reduced Rabi frequencies \cite{wineland_quantum_2003}. For a single ion cooled to the motional ground state, carrier (first blue sideband) $\pi$-times of approximately 13~(15)~\micro s and 5~(23)~\micro s in the axial and radial directions respectively are achieved with 1~mW of light per beam focused to a waist of 40~µm. The laser systems were set up following reference \cite{wilson_750-mw_2011}, except the repumper laser which is based on a frequency doubled DBR diode laser \cite{king_self-injection_2018}. To ensure long-term alignment of the laser beams onto the ion, all 313~nm beams are delivered to the trap through hydrogen-loaded, large mode area optical fibres which have a typical transmission of 50~\% for 1.5~m length \cite{colombe_single-mode_2014,marciniak_towards_2017}.\par
\begin{figure}
	\centering
	\includegraphics[width=1\linewidth]{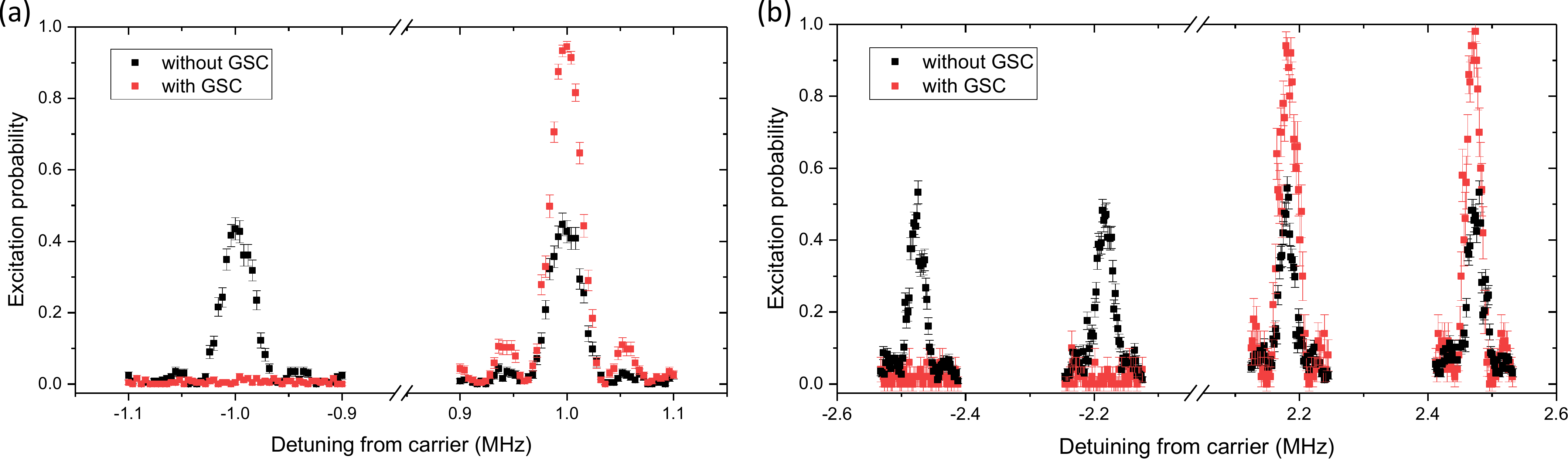}
	\caption{Scan over the red and blue sidebands of the axial motional mode (a) and the radial modes (b) of a $^9$Be$^+$ ion. The black data points were taken with a Doppler-cooled ion, the red data points after ground state cooling. \label{fig:sb_asymmetry}}
\end{figure}

%%%%%%%%%%%%%%%%%%%%%%%%%%%%%%%%%%%%%%%%%%%%%%%%%%%%%%%%%%%%%%%%%%%%%%%%%%%%%%%%
The relatively low secular frequencies somewhat complicate ground-state cooling of the ion, in particular in the axial direction where the Lamb-Dicke parameters for spontaneous emission and stimulated Raman excitation are 0.48 and 0.82, respectively. However, we can exploit this large value for Raman excitation by utilising the high strength of higher-order red sidebands. Several phonons can be removed at once to counteract the effect of recoil heating \cite{wan_efficient_2015,chen_sympathetic_2017,che_efficient_2017}. The cooling process is split into two stages. First, a pre-cooling step consisting of 15 interleaved pulses on each of the fifth, fourth and third red sidebands is applied. This interleaving prevents optical pumping of the ion into "trap" Fock states that have near-zero excitation probability on a particular red sideband. Simulations indicate that this pre-cooling leaves negligible population in Fock states higher than $n=3$. Then, the final cooling step comprises 10 interleaved pulses on each of the second and first order red sidebands. For simplicity, all pulses have the same power and duration, judiciously chosen to match a $\pi$-pulse on the $n=1 \rightarrow n'=0$ transition. In this manner, we are able to reach within 1.8~ms an axial mode ground-state probability of 98~\% with a total of 65 red sideband pulses. The ground state population is calculated using the sideband asymmetry after sideband cooling, see figure \ref{fig:sb_asymmetry}.\par
For radial Raman excitation, the Lamb-Dicke parameter is 0.21, and there are no significant Fock states for which the coupling to red sidebands of a certain order is vanishing. Hence, 20 interleaved sideband pulses on each of the second and first order red sidebands are applied.
% enter GSC stuff

\subsection{Imaging system}
Fluorescence of the Be$^+$ ions on the cooling transition at 313~nm is imaged onto an electron-multiplying CCD camera (Andor iXon3 DU885-KC-VP) and a photon-multiplier tube (PMT, Hamamatsu H10682-210). The fluorescence is split 1:99 between camera and PMT for simultaneous observation of the ion position and its electronic state.\par 
The first imaging lens is a custom bi-aspheric lens (Asphericon) mounted inside the cold stage. It has a free aperture diameter of 22~mm and a working distance of 20~mm. Given its numerical aperture of $\sim$~0.5, it covers 6.9~\% of the total solid angle around the ion. Taking the dipole pattern of circularly polarised spontaneous emission into account reduces the collection efficiency to 5.5~\%.\par 
At cryogenic temperatures, the imaging lens focus must be corrected for thermal contraction. For this purpose the lens is installed in a stress-free copper holder which is in turn mounted onto a piezo-electric translation stage (Attocube Anz101) to adjust the focal distance. To minimise the size of the viewing apertures in the cold stages, the lens relays a near-diffraction-limited image with a magnification of $\times$3 at the exit of a tube analogous to those for the laser beams. This - only slightly magnified - image appears close to a re-entrant room-temperature viewport. It is then further magnified by an air-side lens doublet consisting of an off-the-shelf aspheric lens and a standard plano-convex lens, leading to a total magnification of $\times$22.\par 
With the the laser tuned to resonance and a heavily saturated cooling transition, we observe a fluorescence rate of 340~counts/ms from a single ion, a mere 50~\% of the predicted value including expected losses and the specified PMT quantum efficiency. This could be due to residual misalignment and surface imperfections of the optical elements. For Doppler cooling and state detection we work with an on-resonance count rate of 60~counts/ms, corresponding to 0.2 times the saturation intensity. In this way we prevent saturation broadening that would affect the ion temperature after Doppler cooling and reduce state detection errors due to off-resonant depumping of the $F=1$ state during detection of the ion internal state. At this rate, and inserting a pinhole of 1~mm diameter in front of the PMT for blocking stray light, the total background-count rate (including approximately equal contributions from the PMT dark current and laser scatter) is 2.1~counts/s. Therefore, the signal-to-background ratio reaches a value of 28,500. We discriminate between $\ket{\downarrow}$ and $\ket{\uparrow}$ states (see figure \ref{fig:9Be+}) by counting  fluorescence photons during a fixed time (200~µs) induced by a resonant beam connecting the $^{2}S_{1/2}$ $(F=2)$ manifold to the $P_{3/2}$ manifold and applying a standard thresholding technique \cite{myerson_high-fidelity_2008}. At the given detection efficiency, the state-discrimination fidelity is limited to 98~\% by off-resonant depumping of the dark state $\ket{\uparrow}$ \cite{hemmerling_novel_2012}.

\section{Trap characterisation}

\subsection{Heating rates}
The low mass of $^9$Be$^+$ and the high charge state of a HCI make both species highly susceptible to electric field noise on the trap electrodes, which leads to ion heating and thus higher temperatures after laser cooling. It will also cause systematic shifts of the clock resonance during interrogation if continuous cooling is not applied \cite{rosenband_frequency_2008-1}. In view of the favorable scaling of the heating rate with the secular frequency \cite{turchette_heating_2000}, it would be desirable to operate with the tightest possible confinement of the ion. However, to minimise the RF power dissipated in the cold stages and thus achieve the lowest possible temperature of those parts, the opposite becomes true, and we therefore work with the weakest possible ion confinement which still allows motional ground state cooling. Hence, it is important to minimise electric field noise as far as possible.\par
\begin{figure}
	\centering
	\includegraphics[width=0.8\linewidth]{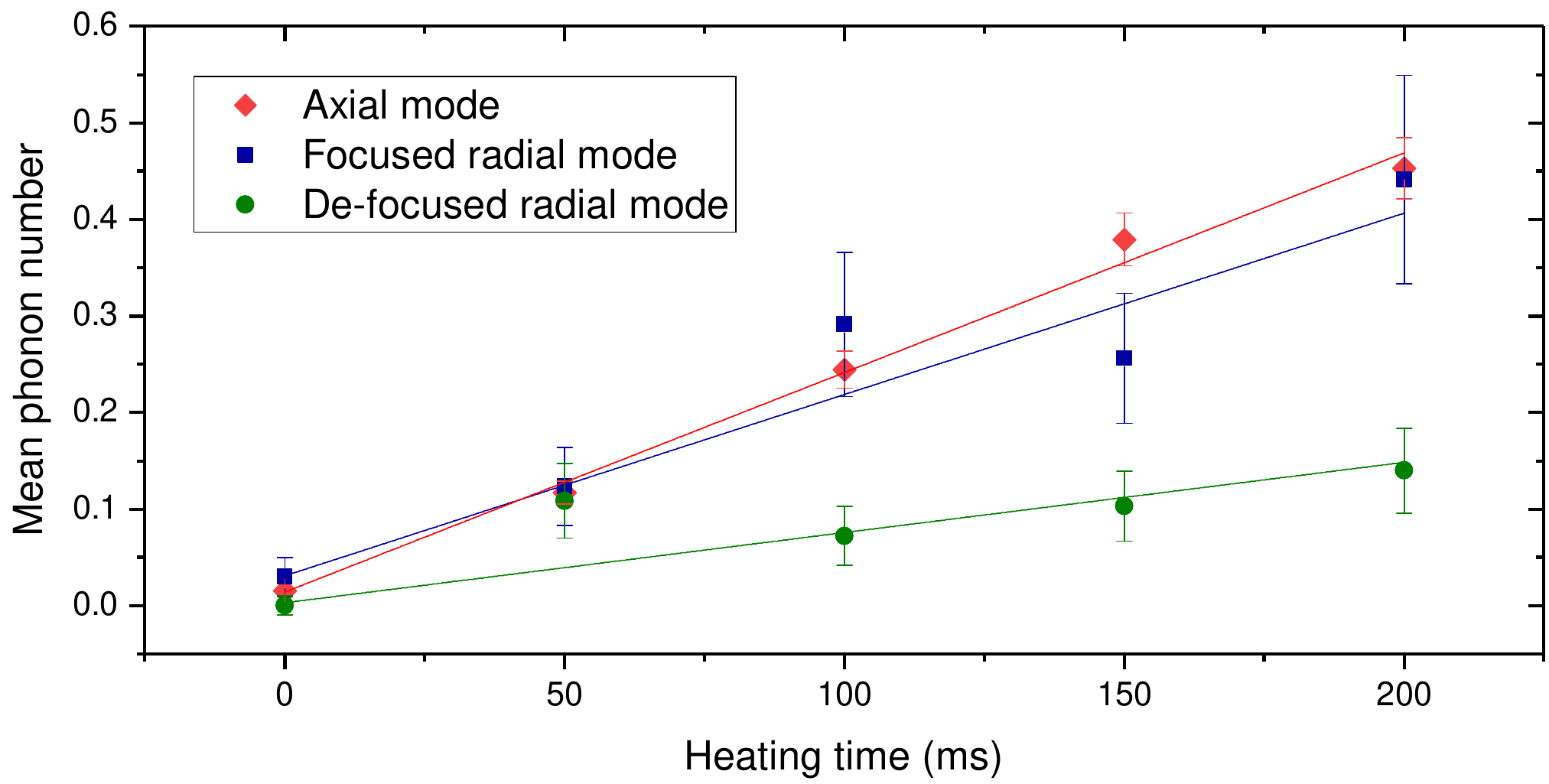}
	\caption{Ion heating on the de-focused (circles) and focused (squares) radial motional mode, as well as the axial mode (diamonds). \label{fig:heating}}
\end{figure}

Measurements of the heating rate were carried out by initialising a single $^9$Be$^+$ ion in the ground state of one or more of its motional modes, then all lasers are turned off to allow the ion to heat freely during periods of up to 200~ms. The final ground-state population was then determined by evaluating the asymmetry of the first order blue and red sidebands of the mode of interest \cite{turchette_heating_2000}. From this, the heating rate was determined to be 2.3(1)~phonons/s in the axial mode at a frequency near 1.0~MHz, and 0.7(2) and 1.9(3)~phonons/s for the two radial modes with frequencies near 2.2~MHz (de-focused) and 2.5~MHz (focused) as shown in figure \ref{fig:heating}. This corresponds to an electric field noise spectral density (in units of $10^{-15}$~V$^2$m$^{-2}$Hz$^{-1}$) of $3.6(2)$ for the axial direction and $2.6(7)$ and $7.1(11)$ for the radial directions. Such values compare well to other cryogenic traps of this size and are between one and two orders of magnitude lower than in traps of similar  designs that are operated at room temperature \cite{brownnutt_ion-trap_2015}. \par 
For measuring the frequency dependence of the heating rate in the radial direction we scan the RF-drive power. The observed heating rate in quanta per second was independent of the secular frequency over the range 1.9 to 3.2~MHz. Noise with an $f^1$ spectrum, or residual uncompensated RF electric field at the position of the ion could explain this effect.\par
Strong filtering of the DC voltages did not affect the anomalous heating rate in the axial direction. Unexpectedly, it was observed that it increased with growing RF drive power, though the radial and axial directions should be mutually independent. In conjunction with the observed independence of the radial heating rate from the RF-trap depth, this indicates that the heating rate in this mode is also limited by RF-drive noise. This couples to the ion through the residual axial RF electric field and causes parametric excitation of the intrinsic micromotion sidebands \cite{sedlacek_method_2018}. A further reduction of the heating rate could therefore be achieved by means of a helical resonator with a higher loaded $Q$~factor thus improving electronic filtering and causing higher trap frequencies for a given level of RF-power dissipation.\par 

\subsection{Micromotion}
If the ion is displaced from the RF nodal line of the trap, or if any on-axis RF field component is induced by electrode alignment inaccuracies, ion motion will be driven at the RF frequency, an effect known as excess micromotion. This is expected to dominate the error budget of an optical frequency standard based on HCI \cite{kozlov_highly_2018}.\par
During curing of the glue holding the various trap parts together, an axial misalignment of approximately 15~\micro m developed between the blade pairs, a value greater than machining tolerances of 5~\micro m for the individual parts.\par
\begin{figure}
	\centering
	\includegraphics[width=0.8\linewidth]{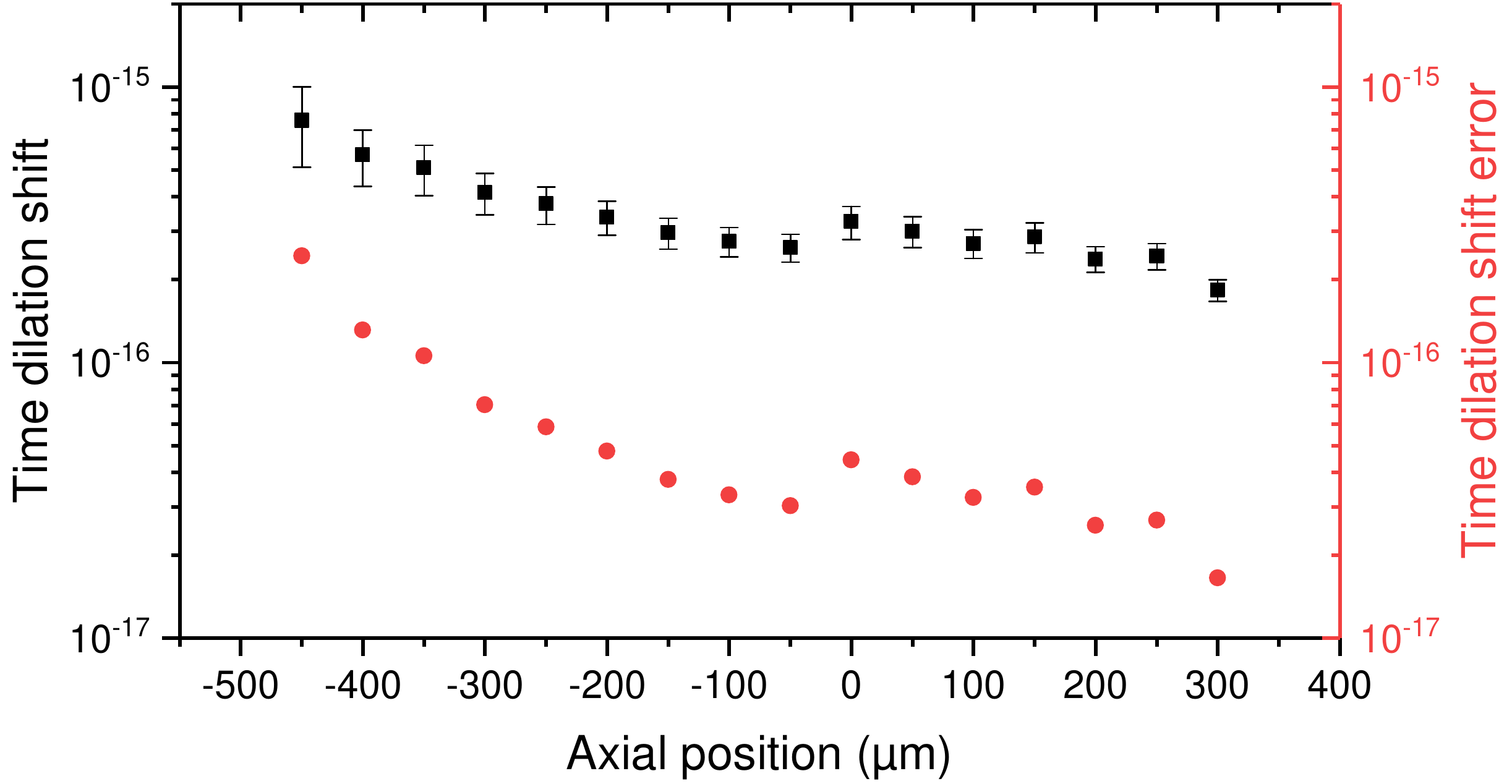}
	\caption{Micromotion-induced fractional time-dilation shift as a function of the ion position along the trap axis. Data were taken with an RF field amplitude corresponding to a radial confinement of $\nu_r=1.5$~MHz for a Be$^+$ ion. \label{fig:MM}}
\end{figure}

 
We measured the axial micromotion with the resolved sideband method \cite{berkeland_minimization_1998,keller_precise_2015} for a radial confinement of $\nu_r=1.5$~MHz for a single $^9$Be$^+$ ion. Figure \ref{fig:MM} shows the axial micromotion-induced time-dilation shift as a function of the axial position in the trap given relative to the ion position for symmetric voltages on the DC electrodes. We sweeped over the entire central segment length under stable trapping conditions. Only a small variation of the micromotion amplitude along the axis was observed, with no zero-crossing of the RF electric field amplitude along the axis. The observed modulation index for the central position was $\beta=1.5$. With a 10~\% error in the determination of the sideband-Rabi frequencies, a fractional time-dilation shift error on the order of $10^{-17}$ is estimated for $^9$Be$^+$.\par 

\subsection{Magnetic field stability}
\begin{figure}	
	\centering
	\includegraphics[width=0.7\textwidth]{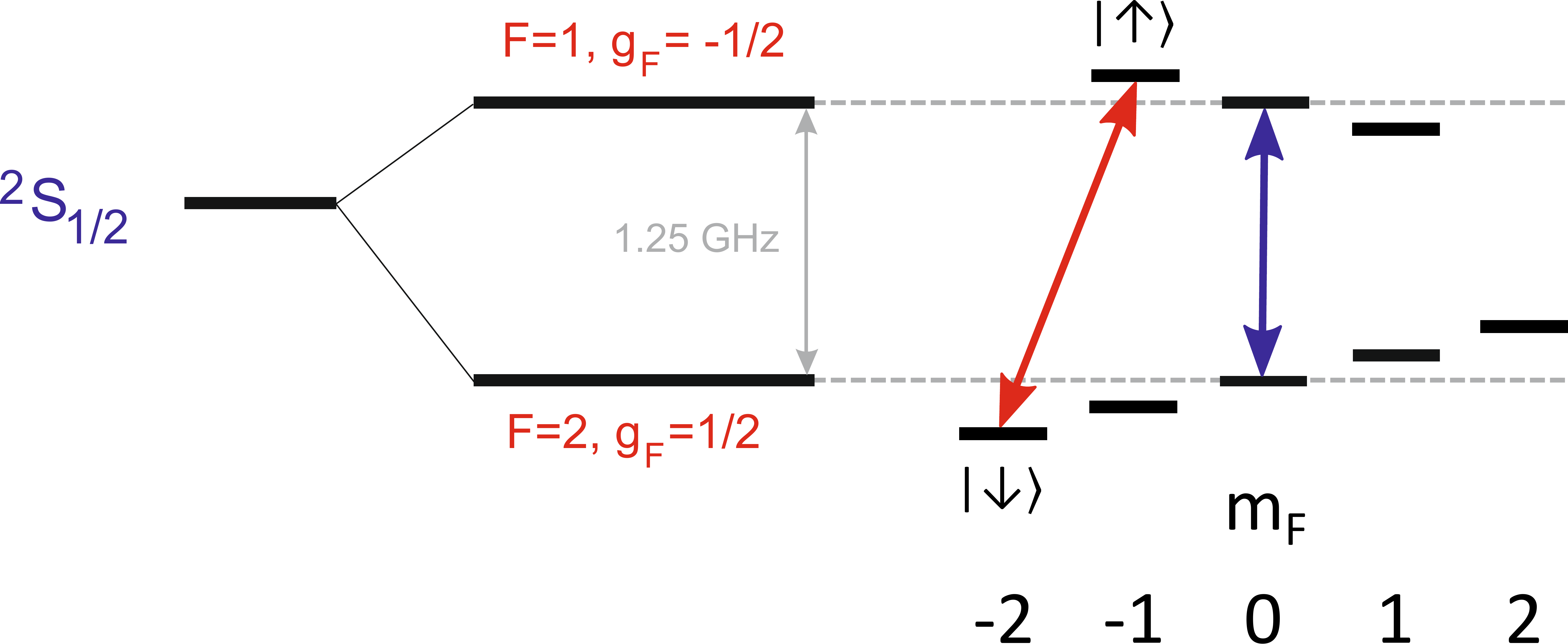} \\
	\caption{Partial term scheme of $^9$Be$^+$ showing the ground state hyperfine and Zeeman structure. Microwave transitions used for setup characterisation are indicated. Red arrow: first-order Zeeman sensitive transition used as the hyperfine qubit, with a shift coefficient of $\Delta\nu/B=21$~Hz/nT. Blue arrow: first-order-insensitive transition.}.
	\label{fig:9Be+}
\end{figure}
Figure \ref{fig:9Be+} shows a partial term scheme of the $^9$Be$^+$ ion, with the hyperfine and Zeeman structure in the  $^2 S_{1/2}$ ground state. Under an external magnetic bias field of about 160~µT, we measure the magnetic field at the position of the ion by means of microwave spectroscopy on the first order magnetic field-sensitive transition $\ket{\downarrow}=\ket{F=2,m_F=-2}$ to $\ket{\uparrow}=\ket{F=1,m_F=-1}$. The linear Zeeman shift of $\Delta\nu/B=(3/2) \mu_B/h\simeq 21$~Hz/nT leads to a transition frequency of 1253.366~MHz. Since the hyperfine constant of Be$^+$ is known to a precision of $10^{-11}$ with $A=-625\,008\,837.048(10)$~Hz \cite{wineland_laser-fluorescence_1983}, we can calculate the linear and quadratic Zeeman shift from the observed splitting between $\ket{\downarrow}$ and $\ket{\uparrow}$, which will subsequently be referred to as the qubit transition.\par 
Despite the lack of magnetic shielding around the vacuum chamber, the short-term stability of the magnetic field at the ion position is improved by the two, at 50~K and 4.5~K highly conductive, copper thermal shields. Alternating magnetic fields induce eddy currents, suppressing magnetic field changes inside the shields \cite{brandl_cryogenic_2016}.  Assuming a low-pass filtering effect of first order, the decay time of these currents gives the corner frequency of the filter function. Although one could expect a double-exponential decay due to the two nested shielding layers, our measurements are compatible with a single low-pass behaviour. 

We measured the ring-down time by observing the step response of the Zeeman splitting for current steps on the magnetic field coils in all three principal axes: $x$, horizontal direction perpendicular to the trap axial direction; $y$, vertical; $z$, trap axial direction. In order to observe the frequency change of the Zeeman splitting we produced an error signal applying two-point sampling with a linear range matched to the observed frequency shift. Several  up and down steps in current were applied, and the exponential rise times averaged.\par 

Symmetry suggests that the shielding effect in both horizontal directions is identical. In the vertical direction the shielding should be better as there are solid horizontal OFHC-copper parts close to and aligned concentrically to the ion position. Shielding in both horizontal directions however suffer from the contact resistances between base, wall and lid elements. Experimental data confirmed these considerations. Table~\ref{tab:step_response} shows the derived time constants and corresponding corner frequencies for the three different principal axis. Within statistics, the values along $x$ and $z$ are the same and about half the one in $y$ direction.\par 
\begin{table}[htb]
	\centering
	\begin{tabular}{ccc}
		\hline
		\hline
		Axis & Time constant (s) & Low-pass corner frequency (Hz) \\
		\hline 
		$x$ & 0.66(18) & 0.24(7) \\
		$y$ & 1.11(17) & 0.14(2) \\
		$z$ & 0.53(10) & 0.30(6) \\
		\hline
		\hline
	\end{tabular}
	\caption{Experimentally determined time constants and corresponding corner frequencies for magnetic field changes along the different axes.\label{tab:step_response}}
\end{table}
Complementary measurements of the shielding factor at higher frequencies were obtained by applying an alternating current to another magnetic coil pair placed outside the main coil set, with the magnitude of the applied field determined using a magnetic flux sensor next to the vacuum chamber. The noise at the ion position was spectroscopically measured using the quantum lock-in amplifier technique \cite{kotler_single-ion_2011}, and indicated a shielding factor of 30-40~dB at frequencies between 60~Hz and 1~kHz.\par 

For evaluation of the low-frequency temporal stability of the magnetic bias field, a closed-loop frequency measurement of the qubit transition was carried out. The microwave power and interaction time were matched to produce a linewidth of 100~Hz with 98~\% contrast using Rabi excitation. With the two-point sampling method, an error signal was created that steered the microwave source. \par 
It was found that, occasionally, changes of up to 50~nT occurred in the field on time scales of several seconds. This was traced down to the motion of an elevator in the building. To compensate for this, two pairs of active magnetic field coils were constructed along the $x$ and $z$ axis, as the quantisation axis lies in the corresponding plane. Field variations are detected using a 3-axis magnetic flux sensor located next to the main vacuum chamber, and feedback with a bandwidth of 1~kHz is then applied to the coil currents for compensation. With this active stabilisation engaged, the elevator no longer affects the field at the ion position to within our resolution of 0.15~nT. \par 
\begin{figure}
	\centering
	\includegraphics[width=1\linewidth]{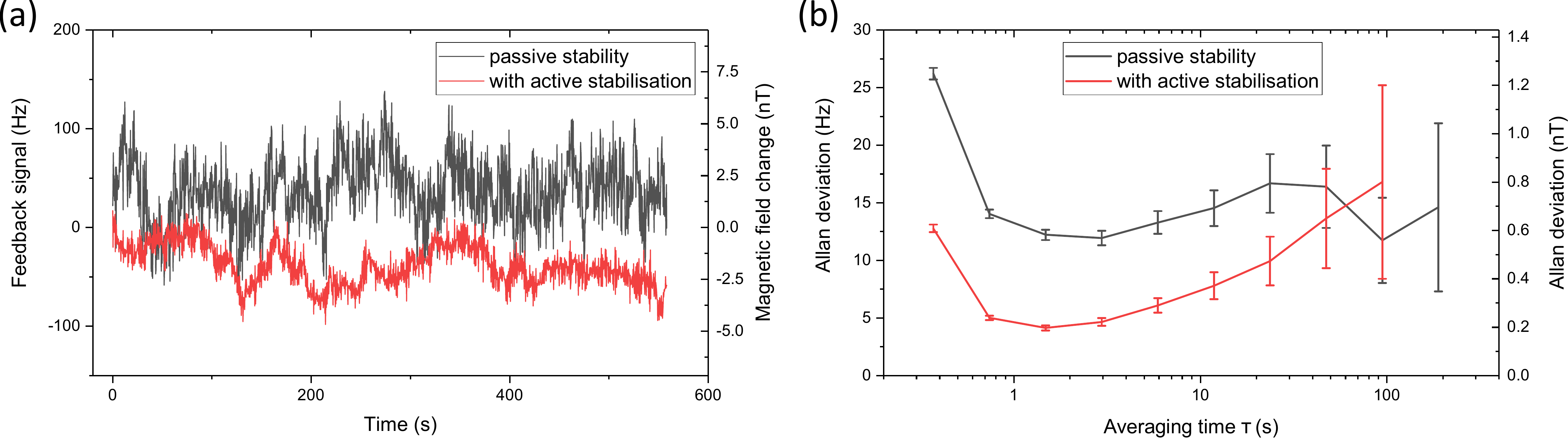}
	\caption{(a) Time trace of the Zeeman splitting of our qubit states relative to the initial value for passive and active magnetic field stability. (b) Allan deviation corresponding to the data from (a) showing the Zeeman-shift stability and the corresponding magnetic field stability. \label{fig:b-field_stability}}
\end{figure}
Figure \ref{fig:b-field_stability} shows a time trace of the frequency feedback necessary to lock to the qubit transitions for the cases of active and passive magnetic field stabilisation. It is apparent from the reduced RMS amplitude of the feedback loop that the active method greatly improves the short-term stability of the field. The Allan deviation shows a passive magnetic field stability of better than 1~nT at time scales from 1 second up to 100 seconds. A fractional magnetic field stability corresponding to 1~nT is below $6\cdot 10^{-6}$ and thus at the limits of commercially available current supplies. Our active stabilisation suppresses fluctuations up to several 10 seconds with an optimum stability of about 200~pT at 1~second. This is close to the limit given by the resolution of the employed magnetic field sensor. Further improvements could result when shielding the vacuum chamber with high-permeability materials to suppress low frequency drifts \cite{altarev_magnetically_2014}.\par 

\subsection{Coherence time}
Decoherence on the qubit transition is caused by energy level shifts due to (a) fluctuating magnetic fields, and (b) fluctuations in power and frequency of the source driving the transition. For microwave excitation, frequency fluctuations of the source can be excluded, as all our radio-frequency devices are referenced to a maser with a stability better than $10^{-12}/\sqrt{\tau}$. Frequency excursions will thus be smaller than the Fourier-limited transition linewidth for all reasonable interrogation times.\par 
In most ion trap experiments, coherence times are limited by technical magnetic field noise at 50~Hz and its harmonics, originating from various electronic devices, as well as switch-mode power supplies radiating at tens of kHz. Commonly reported coherence times are in the 100~µs range for unshielded room temperature experiments \cite{schmidt-kaler_coherence_2003,hemmerling_single_2011}.\par
\begin{figure}
	\centering
	\includegraphics[width=0.8\linewidth]{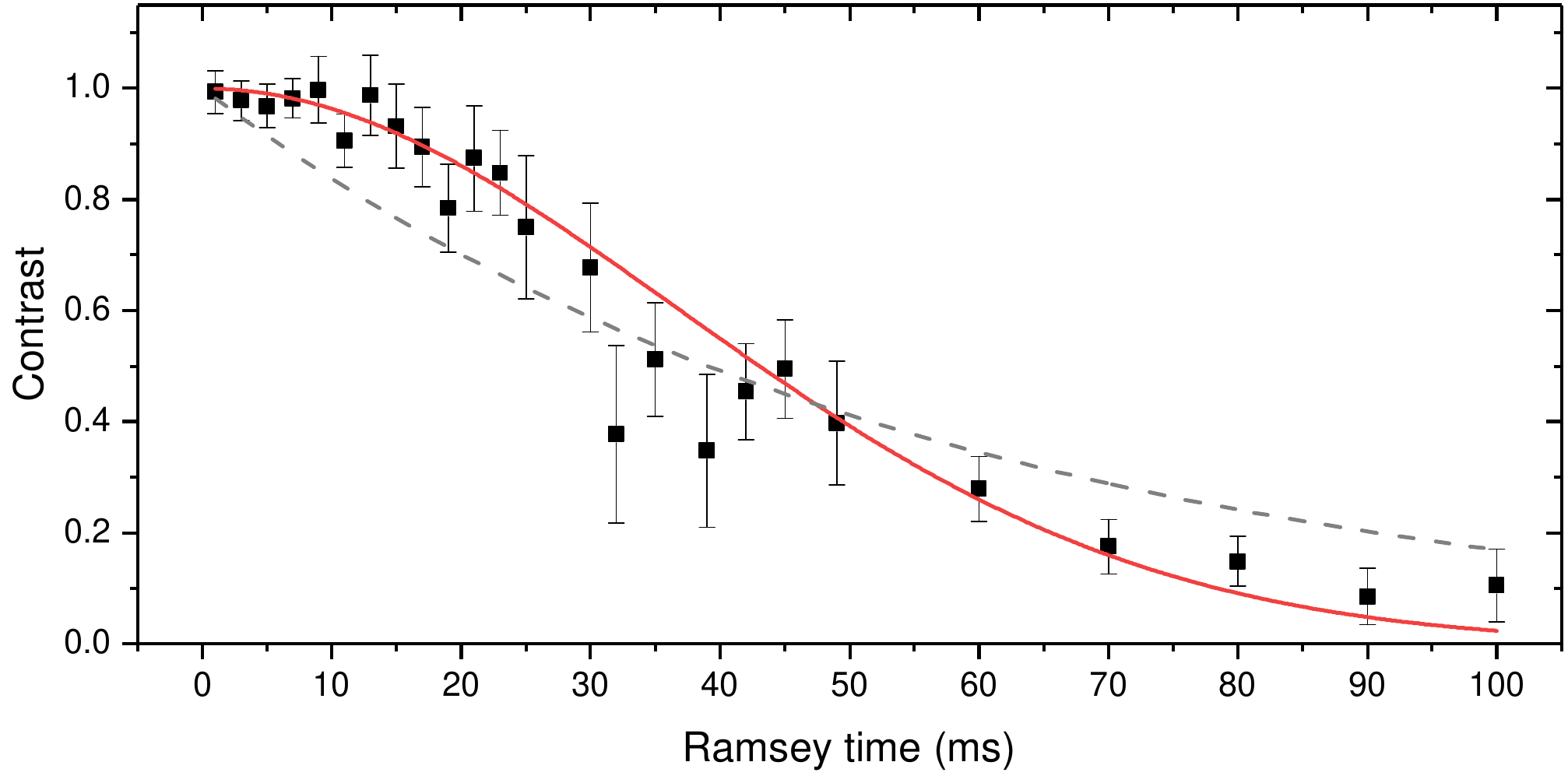}
	\caption{Measured qubit coherence given by the contrast of Ramsey experiments at varying Ramsey delays. The coherence shows a Gaussian envelope (red line) with a $1/\sqrt{e}$ time constant of $\tau=36(1)$~ms. The steep loss of contrast for Ramsey times above $\tau$ is characteristic of our dominant noise process, slow magnetic field drifts. The grey dashed line is a $e^{-t/\tau}$ fit to the data.  \label{fig:coherence_time} }
\end{figure}

 
We measure the coherence time by performing a microwave Ramsey experiment on our qubit transition. With the microwave drive tuned to resonance, the relative phase of the second Ramsey pulse is scanned with respect to the first. The amplitude of the resulting sinusoidal signal indicates the maximum fringe contrast. Measurements for different Ramsey times allow to extract the coherence time, as shown in figure~\ref{fig:coherence_time}. It should be noted that we neither use AC-line triggering nor apply spin-echo sequences \cite{biercuk_optimized_2009,kotler_single-ion_2011,wang_single-qubit_2017} to artificially extend the coherence time, as we want to determine the intrinsic decoherence time scale.
The coherence decay exhibits a Gaussian shape, scaling with $e^{-(t/2\tau)^2}$ if the intrinsic decoherence time is longer than the experimental Ramsey time \cite{monz_quantum_2011,ruster_long-lived_2016}. A fit to the data yields a $1/\sqrt{e}$ coherence time of $\tau=36(1)$~ms. If the intrinsic decoherence is shorter than a single experimental cycle, the contrast falls off exponentially with $e^{-t/\tau}$.  The graph also shows a fit to the data using the exponential decay, which yields $\tau = 56(5)$~ms. However, especially at Ramsey times below 30~ms it is apparent that the data is reproduced more appropriately by the Gaussian envelope, validating the former fit function.\par 
%However, going to slightly longer Ramsey times than 23~ms completely dephases the contrast, indicating the validity of the former fit function. \par 
The measured coherence time is about a factor of three higher than reported for an optical transition in a similar cryogenic system \cite{brandl_cryogenic_2016} and comparable to that of a double layer $\mu$-metal shielded room temperature system, when operating with magnetic field coils inside the shield \cite{ruster_long-lived_2016}.\par
With the measured coherence time $\tau$, the RMS magnetic field fluctuations can be expressed as \cite{ruster_long-lived_2016}
\begin{equation}
\sqrt{\langle \Delta B^2 \rangle} = 2\hbar/(3\mu_B \tau) = 330\textnormal{~pT}\, ,
\end{equation}
taking the magnetic field sensitivity of the transition of $(3/2) \mu_B/h$ into account, where $\mu_B$ is Bohr's magneton. This value is consistent with the measured Allan deviation of the qubit transition frequency at averaging times of 0.5-5~s, corresponding to the measurement time of a single phase scan.

\subsection{Trap-induced ac Zeeman shift}
The oscillating currents on the RF trap electrodes lead to a time-averaged second-order Zeeman shift induced by the radial trapping potential. Due to imbalanced currents in opposing electrodes, this can even be the case for operation at the AC electric field null of the trap \cite{rosenband_frequency_2008-1,gan_oscillating-magnetic-field_2018}. To investigate this effect we performed Ramsey spectroscopy on the $\ket{F=2,m_F=0}$ to $\ket{F=1,m_F=0}$ transition, see figure \ref{fig:9Be+}. This transition is insensitive to first order to magnetic field changes, which enables us to use Ramsey times of 100~ms to resolve the line with a width of 5~Hz. A two-point sampling frequency feedback servo is employed to track the transition frequency. By running interleaved servos on both the magnetic field sensitive transition $\ket{\downarrow}$ to $\ket{\uparrow}$ and the insensitive transition $\ket{F=2,m_F=0}$ to $\ket{F=1,m_F=0}$ we can correct for the second order frequency shift due to drifts in the DC bias field. The first order-sensitive transition is interrogated with a Ramsey time of 10~ms due to its larger frequency fluctuations.\par 
We measure the bias-field-corrected transition frequency $\ket{F=2,m_F=0}$ to $\ket{F=1,m_F=0}$ for three different radial trap depths, corresponding to Be$^+$ secular frequencies of 1.05, 2.42 and 3.68~MHz. Any AC Zeeman shift increases the transition frequency for higher radial secular frequencies. We see no significant shift at the level of our statistics-limited resolution of $\sim$1 Hz. With an AC Zeeman sensitivity of $314$~mHz/µT$^2$, this provides an upper bound of $\langle B^2 \rangle <3$~µT$^2$ for trap-induced magnetic fields. Owing to the strong scaling of the frequency of magnetic dipole transitions with charge state \cite{gillaspy_highly_2001}, the equivalent AC Zeeman sensitivity in HCI is expected to be 3-4 orders of magnitude smaller than in singly charged systems \cite{berengut_highly_2012}.\par
\begin{figure}
	\centering
	\includegraphics[width=0.8\linewidth]{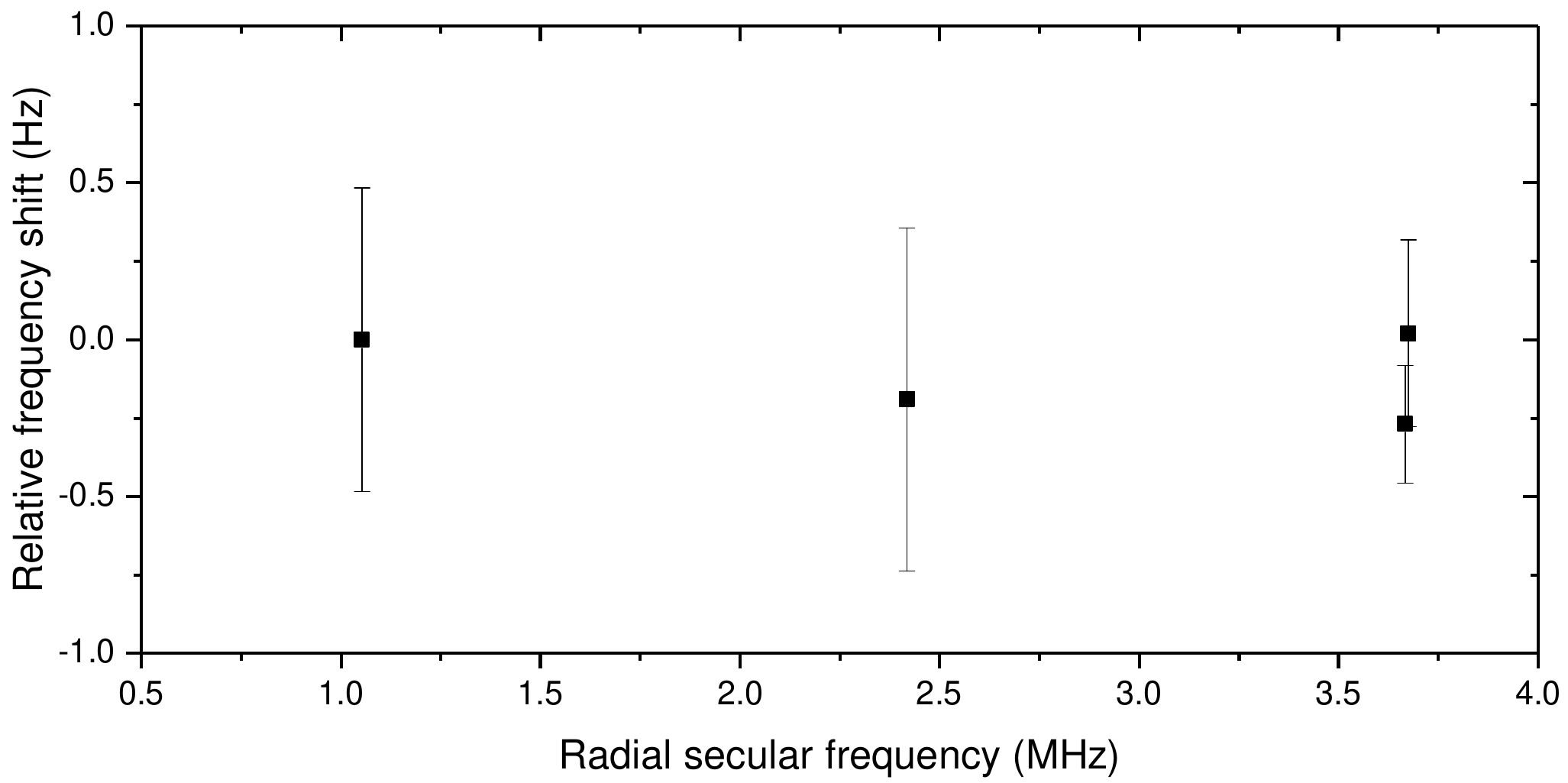}
	\caption{Relative frequency shift of the $\ket{F=2,m_F=0}$ to $\ket{F=1,m_F=0}$ transition as a function of one of the radial secular frequencies. \label{fig:2nd_order_zeeman}}
\end{figure}

\section{Trapping of HCI}\label{HCI}
Highly charged argon ions are extracted in bunches from an EBIT with a mean energy of 700~eV per unit charge, and decelerated to approximately 150~eV per unit charge using a pulse-electrode potential elevator described in reference \cite{schmoger_elektrodynamisches_2013}. Operation of the EBIT and deceleration beamline is described in detail elsewhere \cite{micke_heidelberg_2018,schmoger_deceleration_2015}. \par 
A single charge state (Ar$^{13+}$) is selected by its time of flight for gated injection into the Paul trap. The final deceleration step is performed by raising the trap DC and RF ground to a few volts below the remaining kinetic energy per unit charge. Once the ions have passed the mirror electrode 1 (see figure \ref{fig:trap_rendering}), the electrode is switched to a higher potential, confining the HCI axially, together with the second pair of outer end\-caps. The inner endcaps provide a weak axial confinement for a Be$^+$ Coulomb crystal with 200~mV of applied voltage. Repeated passes and interactions with the Be$^+$ dissipate the kinetic energy of the HCI \cite{gruber_evidence_2001,bussmann_stopping_2006,bussmann_simulating_2007,schmoger_deceleration_2015} until a HCI becomes embedded in the Coulomb crystal, leaving as signature a large dark spot in the otherwise fluorescing crystal. For quantum logic operations, a two-ion crystal of a Be$^+$ ion and the spectroscopy ion is needed. Thus, after co-crystallisation of the HCI, the excess Be$^+$ ions are ejected from the trap by parametric heating using the RF drive.\par

\begin{figure}
	\centering
	\includegraphics[width=1\linewidth]{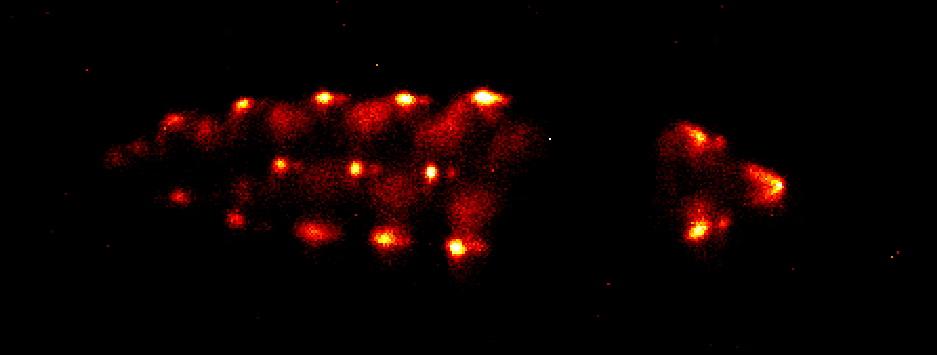}
	\caption{A single Ar$^{13+}$ ion in a Coulomb crystal of several 10 fluorescing Be$^+$ ions. Due to its high charge, the Ar$^{13+}$ displaces several Be$^+$, appearing as a large dark spot in the crystal. \label{fig:HCI}}
\end{figure}

\section{Conclusion}\label{Con}
We have presented a cryogenic ion trap system specifically designed for spectroscopy of single HCI, and characterised the ion trap with respect to heating rates, excess micromotion and magnetic field stability. Ground state cooling of all the normal modes of a single $^9$Be$^+$ ion has been demonstrated outside the Lamb-Dicke regime. With the demonstrated specifications matching the design values, the system is well suited for performing quantum logic spectroscopy not only with HCI, as planned, but also in other species such as molecular ions that would undoubtedly benefit from the extremely low level of BBR, the long coherence times resulting from the active cancellation of external magnetic field fluctuations and the double Faraday shielding by the high-conductivity thermal shields, and the very low density of residual gas particles resulting from the operation near 4\,K. Further additions such as passive magnetic shielding will help improving its advantages.

\ack
The authors gratefully thank Julia Fenske for the design and manufacturing of the electronics required for operation of the trap. We acknowledge the PTB scientific instrumentation department 5.5 headed by Frank L\"offler for their expertise and fabrication of numerous parts, including the cryogenic environment; in particular, we thank Michael M\"uller and Stephan Metschke. We thank the PTB surface technology laboratory and Manuel Stompe from IMPT Hannover for their technical help with manufacturing of the ion trap. We thank the MPIK mechanical apprentice workshop headed by Stefan Flicker for fabrication of parts. Financial support was provided by Physikalisch-Technische Bundesanstalt, Max-Planck-Gesellschaft zur F\"orderung der Wissenschaften e.~V.~, and Deutsche Forschungsgemeinschaft through SCHM2678/5-1, and Collaborative Research Centers SFB~1227 (DQ-mat), projects A01 and B05, as well as SFB~1225 (ISOQUANT), project B01. SAK acknowledges support by the Alexander von Humboldt Foundation.

%\bibliographystyle{unsrt}
%\bibliography{2018_CryoTrap2}

\end{document}